\documentclass[useAMS,usenatbib,usegraphicx]{mn2e}
\usepackage{times}

\title[]{Study of Isotopic Fractions and Abundances of the Neutron-capture Elements in HD175305}
\author[Jiang ZHANG $^{1,2}$, Wenyuan CUI $^{1}$ and Bo Zhang]{Jiang Zhang$^{1,2}$, Wenyuan
Cui$^{1}$ and Bo Zhang$^{1}$\thanks{Corresponding author.
Email-address: zhangbo@mail.hebtu.edu.cn (Bo Zhang)}\\
$^{1}$Department of Physics, Hebei Normal University, 113 Yuhua Dong
Road, Shijiazhuang 050016, P.R.China\\  \hspace{ 1mm } Hebei
Advanced Thin Films
Laboratory, shijiazhuang 050016\\
$^{2}$School of Mathematics and Science, Shijiazhuang University of
Economics, Shijiazhuang 050016, P.R.China}
\begin{document}

\date{Received...; in original form...}

\pagerange{\pageref{firstpage}--\pageref{lastpage}} \pubyear{2009}

\maketitle

\label{firstpage}

\begin{abstract}
The chemical abundances of the metal-poor stars are excellent
information for setting new constraints on models of galaxy chemical
evolution at low metallicity. In this paper we present an attempt to
fit the elemental abundances observed in the bright, metal-poor
giant HD 175305 and derive isotopic fractions using a parametric
model. The observed abundances can be well matched by the combined
contributions from s- and r-processes material. The component
coefficients of the r- and s-processes are $C_1=3.220$ and
$C_3=1.134$, respectively. The Sm isotopic fraction in this star
where the observed neutron-capture elements were produced is
predicted to be $f_{152+154}=0.582$, which suggests that even though
the r-process is dominantly responsible for synthesis of the
neutron-capture elements in early galaxy, the onset of the s-process
has already occurred at this metallicity of $[Fe/H]=-1.6$.
\end{abstract}


\section{Introduction}
The abundances of neutron-capture elements in metal-poor halo stars
are now providing important clues to the chemical evolution and
early history of the Galaxy. These elements are predominantly
produced through successive neutron-capture reactions in two
processes known as the rapid (r-) and the slow (s-) process
(Burbidge et al. 1957). The r-process is usually associated with the
explosive environment of Type II supernovae (SNeII), although this
astrophysical site has not been fully confirmed yet (Cowan et al.
1991, Sneden et al. 2008). The s-process is further divided into two
categories: the weak s-component and the main s-component. It is
well known that massive stars are the sites of the so-called weak
component of s-process nucleosynthesis, which is mainly responsible
for the production of lighter elements (Sr, Y, and Zr) (Lamb et al.
1977; Raiteri et al. 1991, 1993; The et al. 2000). The main
s-process component is responsible for the production of the heavier
s-process elements (heavier than Ba). Observational evidence and
theoretical studies have identified the main s-process site in low-
to intermediate-mass ($\approx1.3-8M_{\odot}$) stars in the
asymptotic giant branch (AGB) (Busso et al. 1999). The observations
in the ultra-metal-poor halo star CS 22892-052 revealed that the
abundances of the heavier ($Z\geq56$) stable neutron-capture
elements are in remarkable agreement with the scaled solar system
r-process pattern, called as ``main r-process stars" (Sneden et al.
2003). More recent work (Sneden et al. 2009), utilizing updated
experimental atomic data to determine more accurate abundances, has
confirmed this agreement. However, some very metal-poor stars, such
as HD 122563 and HD 88609, have high abundance ratios of light
neutron-capture elements Sr, Y, and Zr, while their heavy ones
(e.g., Eu) are very deficient (Honda et al. 2007 ). Such objects
possibly record the abundance patterns produced by another process
and are called as a ``weak r-process" star (Izutani, Umeda, \&
Tominaga 2009). Thus, two separate r-processes are required to
explain the elemental abundances in metal-poor stars .

The solar system heavy-nuclide abundance distribution has been
detected in most detail, so it is usually regarded as a standard
pattern, which is usually compared with the elemental abundances in
metal-poor stars. Cowan et al. (1996) measured abundances of the
r-process peak elements Os and Pt in the metal-poor halo star HD
126238 ($[Fe/H]=-1.7$) and found that neither the solar pure
r-process nor the total solar abundance can fit all of the observed
abundances of the heavy elements in this star. They obtained a
better fit by taking some fraction of s-process abundances and
adding it to the solar r-process distribution, which suggested that
the onset of the s-process has already occurred at this metallicity.
As the abundances of the metal-poor halo stars can be used as a
probe of the conditions that existed early in the history of the
Galaxy, it is important and necessary to determine the relative
contributions from the individual neutron processes to the
abundances of heavy elements in metal-poor stars.

 Additional evidence of the earliest Galactic
neutron-capture nucleosynthesis can be obtained by isotopic
abundance determinations of metal-poor stars (Sneden et al. 2002;
Lambert \& Allende Prieto 2002; Aoki et al. 2003; Mashonkina \& Zhao
2006; Kratz et al. 2007). Since the isotopic abundances can be
directly used to assess the relative s- and r-process contributions
without the smearing effect of multiple isotopes, they should be
more fundamental indicators of neutron-capture nucleosynthesis.
 Recently, Roederer et al. (2008) measured
abundances and isotopic fractions of neutron-capture elements in the
metal-poor star HD 175305 ($[Fe/H]=-1.6$) and concluded that the
observed Sm isotopic fraction is consistent with r-process origin.
However, they found that a scaled solar system r-process pattern
cannot fit not only Ba and Eu abundance but also the Sm isotopic
fraction measured from the line at 4604.17 \AA. Clearly, the close
examination of elemental abundances and isotopic fraction in this
object is very important for a good understanding of the abundance
pattern of neutron-capture element in the metal-poor stars. These
reasons motivated us to investigate the ratio of s- and r- processes
in this star. It is interesting to adopt the parametric model for
metal-poor stars to study the relative contributions from the s- and
r- processes that could reproduce the observed abundance pattern
found in this star. The parametric model of metal-poor stars is
described in Sect.\ 2. The calculated results are presented in
Sect.\ 3 in which we also discuss the characteristics of the s- and
r-process nucleosynthesis. Conclusions are given in Sect.\ 4.

\section{Parametric model of Metal-Poor Stars}

  Generally, the solar system nuclide abundance distribution
serves as the standard pattern or basis when investigating the
abundance distribution of neutron-capture elements. However, the
observational studies have shown that the abundance patterns of
heavy elements in population II stars differ obviously from the
solar (Spite \& Spite 1978; Gratton et al. 1994), and the same is
true even for population I stars (Woolf et al. 1995). Based on
this idea, Zhang et al. (1999) presented a model for calculating
the abundances of heavy elements in metal-poor stars. Their
results showed that the observed abundances of heavy elements of
the sample of Gratton \& Sneden (1994) can be well matched by an
abundance mix of the solar r- and s-processes, especially for
elements heavier than Ba.

We explored the origin of the heavy neutron-capture elements
($Z\ge56$) in HD 175305 by comparing the observed abundances with
predicted r- and s-process contributions. Since the weak r-process
that is responsible for the elements in the Sr-Ag range would have
to provide negligible contributions to Ba while the main r-process
is known to produce the heavy elements from Ba and beyond (Montes et
al. 2007), we ignore the contributions of weak r-process to heavier
elements ($Z\geq56$) in our calculations. The $i$th element's
abundance in the star can be calculated as follows (Zhang et al.
1999):
\begin{equation}
N_{i}(Z)=(C_{1}N_{i,\ r}+C_{3}N_{i,\ s,main})10^{[Fe/H]},
\end{equation}
where $N_{i,\ r}$ and $N_{i,\ s,main}$ are the abundances of the
$i$th element produced by the r-process and the main s-process in
solar material, respectively. $C_1$ and $C_3$ are the component
coefficients that correspond to the relative contributions from
the r-process and the main s-process to the element abundances in
metal-poor stars, respectively. Here we choose the solar component
coefficients as a standard and assume all of them are equal to 1,
while the solar r- and s-process abundances are adopted from
Arlandini et al. (1999).The La value in solar system is updated
from O'Brien et al. (2003), and the Hf value from Lawler et al.
(2007).

Our goal is to find the parameters which best characterize the
observed data. The reduced $\chi^2$ is defined
\begin{equation}
 \chi^2 = \sum \frac{(N_{obs}-N_{cal})^2}{(\bigtriangleup N_{obs})^2(K-K_{free})}, \nonumber \\
\end{equation}
where$\Delta N_{obs}$ is the uncertainty on the observed
abundance, $K$ is the number of elements used in the fit, and
$K_{free}$ is the number of free parameters varied in the fit. We
then took linear combinations of the solar s- and r-process
abundances (adopting $[Fe/H]=-1.6$) exploring the full range of
parameter space (i.e., considering values from 0.0001 to 5.0000
times the solar s- and r-process abundances). For each combination
of s- and r-process abundances,we compared the predicted abundance
ratios with the observed abundance ratios. The best fit between
theoretical predictions and observational data correspond to the
smallest value of $\chi^2$. For a good fit the reduced $\chi^2$
should be a number that is not very far from 1, which it means
that $\chi^2$ should be of order unity. Based on equation (1), we
carry out the calculation to fit the abundance profile observed in
HD 175305, in order to look for the minimum $\chi^2$ in the two
parameter space formed by $C_{1}$, and $C_{3}$.

\section{Results and Discussion}

Using the observed data in HD 175305 (Roederer et al. 2008), the
model parameters can be obtained. The s- and r-processes component
coefficients of HD 175305 are $C_1 = 3.220$, $C_3 =1.134$, which
implies that this star is not r-only star. Taking the values of
$C_1$, and $C_3$ into Eq. (1), the abundances of all the
neutron-capture elements in HD 175305 are obtained as listed in
Table 1. Columns (3), (4), and (5) of Table 1 give the abundances of
the r-process and main s-components, and the total abundances in
this halo star HD 175305, respectively. In columns (6) and (7), we
list the r- and s-process fractions to the total abundances. These
numbers are useful in understanding the relative dominance of two
processes for a given neutron-capture element in HD 175305.

\begin{table*}
 \centering
 \begin{minipage}{140mm}
  \caption{The r- and s-processes elemental abundances of HD
175305}
  \begin{tabular}{lccccclrc}
  \hline
Element&Z&$N_r$&$N_{s,main}$&$N_{total}$&$r-{fraction}$&$s-{fraction}$\\
 \hline
Ba  & 56    & 6.89E-02  & 9.33E-02  & 1.62E-01  & 0.425 & 0.575 \\
La  & 57    & 9.71E-03  & 9.29E-03  & 1.90E-02  & 0.511 & 0.489 \\
Ce  & 58    & 2.16E-02  & 2.46E-02  & 4.62E-02  & 0.467 & 0.533 \\
Pr  & 59    & 6.93E-03  & 2.32E-03  & 9.25E-03  & 0.750 & 0.250 \\
Nd  & 60    & 2.46E-02  & 1.31E-02  & 3.77E-02  & 0.653 & 0.347 \\
Sm  & 62    & 1.41E-02  & 2.17E-03  & 1.63E-02  & 0.866 & 0.134 \\
Eu  & 63    & 7.42E-03  & 1.60E-04  & 7.58E-03  & 0.979 & 0.021 \\
Gd  & 64    & 2.26E-02  & 1.44E-03  & 2.40E-02  & 0.940 & 0.060 \\
Tb  & 65    & 4.52E-03  & 1.24E-04  & 4.65E-03  & 0.973 & 0.027 \\
Dy  & 66    & 2.71E-02  & 1.66E-03  & 2.87E-02  & 0.942 & 0.058 \\
Ho  & 67    & 6.63E-03  & 1.98E-04  & 6.83E-03  & 0.971 & 0.029 \\
Er  & 68    & 1.68E-02  & 1.21E-03  & 1.80E-02  & 0.933 & 0.067 \\
Tm  & 69    & 2.65E-03  & 1.43E-04  & 2.80E-03  & 0.949 & 0.051 \\
Yb  & 70    & 1.35E-02  & 2.09E-03  & 1.56E-02  & 0.866 & 0.134 \\
Lu  & 71    & 2.38E-03  & 1.80E-04  & 2.56E-03  & 0.930 & 0.070 \\
Hf  & 72    & 8.01E-03  & 2.16E-03  & 1.02E-02  & 0.787 & 0.213 \\
Ta  & 73    & 9.87E-04  & 2.44E-04  & 1.23E-03  & 0.802 & 0.198 \\
W   & 74    & 4.74E-03  & 2.11E-03  & 6.86E-03  & 0.692 & 0.308 \\
Re  & 75    & 4.00E-03  & 1.38E-04  & 4.14E-03  & 0.967 & 0.033 \\
Os  & 76    & 4.91E-02  & 1.80E-03  & 5.09E-02  & 0.965 & 0.035 \\
Ir  & 77    & 5.27E-02  & 2.59E-04  & 5.30E-02  & 0.995 & 0.005 \\
Pt  & 78    & 1.03E-01  & 1.95E-03  & 1.05E-01  & 0.981 & 0.019 \\
Au  & 79    & 1.42E-02  & 3.10E-04  & 1.45E-02  & 0.979 & 0.021 \\
Hg  & 80    & 1.07E-02  & 4.94E-03  & 1.56E-02  & 0.684 & 0.316 \\
Ti  & 81    & 3.60E-03  & 3.97E-03  & 7.57E-03  & 0.475 & 0.525 \\
Pb  & 82    & 5.03E-02  & 6.50E-02  & 1.15E-01  & 0.436 & 0.564 \\
Bi  & 83    & 1.11E-02  & 2.01E-04  & 1.13E-02  & 0.982 & 0.018 \\

\hline

\end{tabular}
\\ NOTE.--- log$\varepsilon$(El)=logN(El)+1.54
\end{minipage}
\end{table*}

The Ba and Eu abundances are useful for investigating contributions
of the s- and r-processes corresponding to those in metal-poor
stars. In the Sun, the elemental abundances of Ba and Eu consist of
significantly different combinations of s- and r-process isotope
contributions, with s:r ratios for Ba and Eu of $80:20$ and $6:94$,
respectively (Arlandini et al. 1999). We explored the contributions
of s- and r-processes for these two elements in HD 175305. In Table
1 we display the s-process fractions calculated from equation (1)
for the sample star. Clearly, for the star listed in Table 1, the
s:r ratios for Ba and Eu are $58:42$ and $2:98$, which are smaller
than the ratios in the solar system, but the contributions from the
s-process is not negligible at all, especially for Ba. Because the
predictions are based on the observed abundances of the
neutron-capture elements and the solar abundances, both the
measurement errors of the stellar abundances and uncertainties of
the solar abundances must be involved in the predictions. The errors
of predictions resulting from the measurement errors of the ratios
$[el_{i}/Fe]$ can be calculated by our model; the mean error is
about $\pm0.15$ dex in log$\varepsilon$. In Fig. 1, the solid line
is the predicted abundances curve, and the two dotted lines show the
error range of the predictions. In order to facilitate the
comparisons of the theoretical abundances with the observations, the
heavy element observed abundances are marked by filled circles. The
figure shows that the model predictions for abundances starting with
Ba fit well the observed data for HD 175305. The curves produced by
the model are consistent with the observed abundances within the
error limits (i.e.,$\chi^2 =0.774$). The agreement of the model
results with the observations provides strong support for the
validity of the parametric model.

\begin{figure*}
 \centering
 \includegraphics[width=0.88\textwidth,height=0.5\textheight]{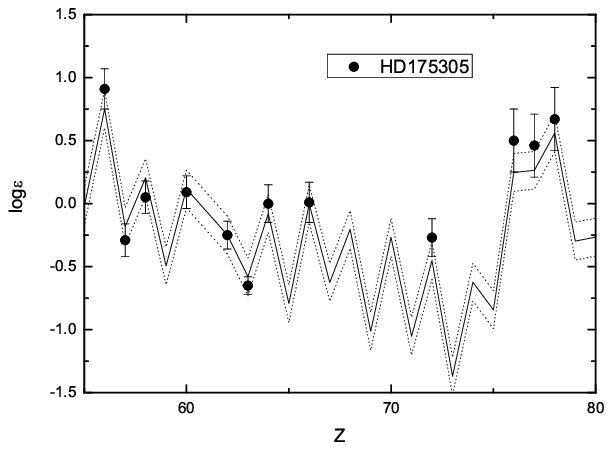}
\caption{Comparison of the observed (Fulbright 2000, Cowan et al.
2005, Lawler et al. 2007) heavy neutron-capture elemental abundances
for HD 175305 with the elemental abundances calculated by this
model. The black filled circles with appropriate error bars denote
the observed element abundances and the solid lines represent
predictions. The two dotted lines are the upper limits and lower
limits of the model calculations, respectively.}

\end{figure*}

\begin{figure*}
 \centering
 \includegraphics[width=0.88
 \textwidth,height=0.5\textheight]{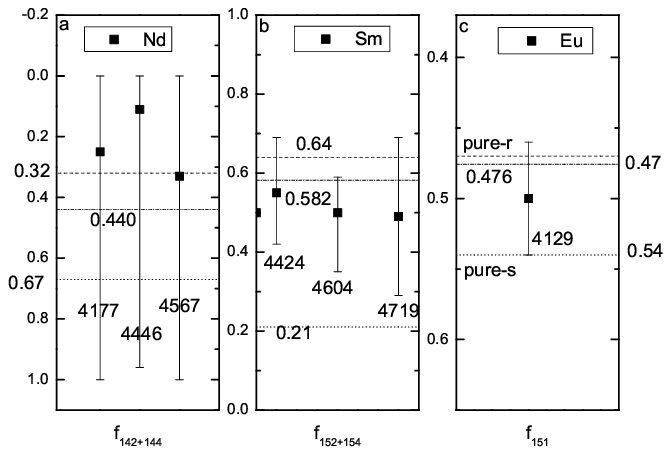}
\caption{Comparison of the isotopic fraction with predicted isotopic
abundances. A summary of measurements the isotopic fraction is
indicated for three species. The isotopic fractions of Nd, Sm, and
Eu as measured quantitatively in HD175305, after Roederer et al.
2008. Squares represent measurements in HD 175305 and the dash
dotted lines represent predictions, while the dotted line and dashed
lines in each panel represent the pure-s and pure-r-process,
respectively, given by Arlandini et al. (1999).}
\end{figure*}

It was possible to isolate the contributions corresponding to the s-
and r-processes for isotopes by our parametric model, which provide
the knowledge of the isotopic composition of the neutron-capture
elements. In the recent studies, the usefulness of Sm, as a
diagnostic tool, is presented to indicate the relative contribution
of r- versus s-process material (Lundqvist et al. 2007). Roederer et
al. (2008) measured isotopic fractions of three neutron-capture
elements including Sm in the metal-poor star HD 175305. The Nd, Sm
and Eu isotopic fractions are defined by
\begin{eqnarray}
f_{142+144}&=&\frac{N(^{142}Nd)+N(^{144}Nd)}{N(Nd)}\nonumber \\
f_{152+154}&=&\frac{N(^{152}Sm) + N(^{154}Sm)}{N(Sm)} \nonumber \\
 f_{151}&=&\frac{N(^{151}Eu)}{N(Eu)}
\end{eqnarray}
In the Sun, the elemental abundance of Sm consists of s- and
r-process isotope contributions, with Sm isotopic fraction
$f_{152+154}$ for pure s- and r-processes of 0.21 and 0.64,
respectively (Arlandini et al. 1999). In HD 175305, the measured Sm
isotopic fraction $f_{152+154} = 0.51_{-0.08}^{+0.08}$ (Reoderer et
al. 2008) doesn't suggest a pure r-process origin. The calculated Sm
isotopic fraction is $f_{152+154}=0.582^{+0.000}_{-0.003}$. In
Figure 2, we summarize the Nd, Sm and Eu isotopic fractions and the
difference between values for pure s- and r-processes. The dash
dotted lines are the predicted isotopic fraction of Nd, Sm and Eu,
and then the dotted and dashed lines show the isotopic fractions of
Nd, Sm and Eu for pure s- and r-processes, respectively. The
observed isotopic fractions of Nd, Sm and Eu are marked by black
filled squares. The figure shows that the model predictions for
isotopic fractions of Nd, Sm and Eu are in good agreement with the
observed data for HD 175305 too. Our calculations agree well with
not only abundances of neutron-capture elements, but also isotopic
fractions, especially the Sm isotopic fraction measured from the
line at 4604.17 \AA. The results mean that most of the
neutron-capture elements are synthesized by a combination of the s-
and r-processes in this metal-poor star, while the contributions
from the s-process is not negligible at all, especially for Ba.
Unfortunately, the Nd isotopic wavelength splits are simply too
small for the lines that were available so that very large error
bars attend the $f_{142+144}$ estimates, as shown in Figure 1a. The
Nd isotopic fraction, $f_{142+144}=0.440$ derived from our
calculation is not suggestive of an pure r-process origin. Our Eu
isotopic fraction, $f_{151}=0.476$ is shown to be a reasonable match
for the measurements derived from only the 4129 \AA, which is in
good agreement with the solar photospheric ratio $0.50\pm0.07$
(Lawler et al. 2001) and the precise Eu meteoritic ratio of
$f_{151}$ = 0.478 (Lodders 2003).

We also discuss the dependence of the calculated abundances
log$\varepsilon$(Ba) and log$\varepsilon$(Eu) on the component
coefficients and uncertainty of the parameters. In the parametric
model, Ba, La and Ce, which are produced dominantly by s-process in
the solar system, are at the same degree of rank/standing. Because
log$\varepsilon$(Ba) is more sensitive to the s-process component
coefficient $C_{3}$ among three elements (i.e., Ba, La and Ce), we
choose Ba as the representative element. Figures 3 and 4 show the
calculated log$\varepsilon$(Ba) and log$\varepsilon$(Eu), as a
function of the the r-process component coefficient $C_1$ in a model
with a fixed $C_3= 1.134$ and as a function of the main s-process
component coefficient $C_3$ with a fixed $C_1=3.220$, respectively.
These are compared with the observed abundances from HD 175305.
There is only one region of $C_1$ in Figure 3,
$C_1=3.220^{+0.003}_{-0.001}$(1$\sigma$), in which both the observed
abundances log$\varepsilon$(Ba) and log$\varepsilon$(Eu) can be
accounted for within the error limits. The bottom panel in Figure 3
displays the $\chi^2 $ value calculated in our model, and there is a
minimum, with $\chi^2 =0.774$, at $C_1=3.220$ with a 1$\sigma$ bar.
It is clear from Figure 3 that log$\varepsilon$(Eu) is very
sensitive to the r-process component coefficient $C_1$. From Figure
4 we can see that the abundance log$\varepsilon$(Eu) is insensitive
to $C_3$ and allows for a wider range $0 < C_3< 1.196$ while
log$\varepsilon$(Ba) is sensitive to $C_3$, $1.133<C_3 < 3.282$.
There is only one region of $C_3$ in Figure 4,
$C_3=1.134^{+0.062}_{-0.001}$(1$\sigma$), in which both the observed
abundances log$\varepsilon$(Ba) and log$\varepsilon$(Eu) can be
accounted for within the error limits.

It is interesting to investigate the component coefficients of
pure r-process stars. The abundances in pure r-process star can be
fit well by the solar-system r-only values ($Z\geq56$) (Sneden et
al. 2008). This means that our model can be used to pure r-process
stars. As an example, for pure r-process star CS 22892-052, the
component coefficient of r-process is about $C_{1}\sim45$ and the
component coefficient of s-process is about $C_{3}\sim0$, which is
consistent with the meaning of pure r-process star.

\begin{figure*}
 \centering
 \includegraphics[width=0.88
 \textwidth,height=0.5\textheight]{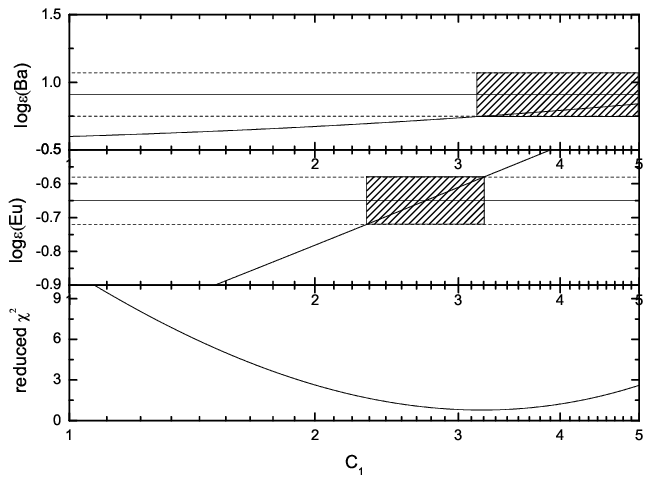}
\caption{Abundance ratios log$\varepsilon$(Ba) (top),
log$\varepsilon$(Eu) (middle), and reduced $\chi^2$ (bottom), as a
function of the r-process component coefficient, $C_1$, in a model
with the s-process component coefficient $C_3=1.134$. Solid curves
refer to the theoretical results, and dashed horizontal lines refer
to the observational results with errors expressed by dotted lines.
The shaded area illustrates the allowed region for the theoretical
model. }
\end{figure*}
\begin{figure*}
 \centering
 \includegraphics[width=0.88
 \textwidth,height=0.5\textheight]{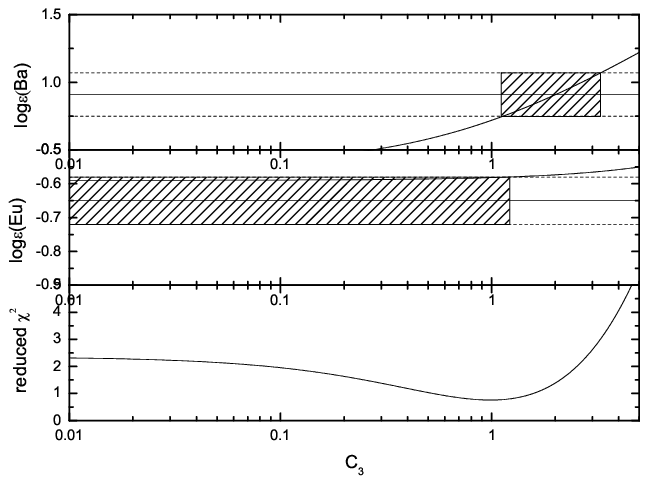}
\caption{Same as those in Fig.\ 3, but as a function of the $C_3$.}
\end{figure*}

\section{ Conclusions}
We have studied the neutron-capture element abundances of HD 175305
in parametric model which includes the mixing of s- and r-processes.
The component coefficients $C_{1}$ and $C_{3}$ contain some
important information for the chemical evolution of the Galaxy. We
find from our results that the relative contribution from the
individual neutron-capture process to the heavy element abundances
in HD 175305 are not similar to that in the solar system.
$C_{1}=3.220$ means that, except for the effect of metallicity, the
contributions from the r-process to the heavy element abundances in
this star is larger than that in the solar system. Moreover,
$C_{1}>C_{3}$ means that the productions of the main r-process are
more remarkable than those of the main s-process in the star. The
trends of the r-process and the main s-process productions in the
early galaxy are obviously different. According to the theory of
stellar evolution, the r-process nuclei are produced during the
stage of massive supernova explosions, which can be synthesized in
significant amounts in the early Galaxy, therefore the r-process
contributions to the heavy elements are dominant at low
metallicities. In contrast, the main s-process occurs during the
phase of thermal pulsation in intermediate and low mass AGB stars,
and the production of the s-process elements needs Fe-seed source,
so its contributions to the heavy elements are very small in the
early Galaxy. Thus, the present study is confined to $[Fe/H]\sim-
1.6$ where AGB stars would contribute in small amounts to the
interstellar medium, which are consistent with the theory of
galactic chemical evolution (Travaglio et al.1999).

We have also calculated isotopic fractions of Eu, Sm, and Nd in the
metal-poor star HD 175305. The isotopic fractions in HD 175305 are
interpreted by isotopic mixes for pure s- and r-processes. In HD
175305, our Sm isotopic fraction
$f_{152+154}=0.582^{+0.000}_{-0.003}$, which is consistent with the
measurements from the 4424 \AA, 4604 \AA and 4719 \AA lines,
suggests that even though the r-process is dominantly responsible
for synthesis of the neutron-capture element in early galaxy, the
onset of the s-process has already occurred at this metallicity of
$[Fe/H]=-1.6$. $C_3=1.134$ means that, except for the effect of
metallicity, the contributions from the main s-process to the heavy
element abundances in this star have reach solar system values. The
isotopic abundance fractions, derived from the our model, are in
excellent agreement with the observed values, which means that the
results have extended the agreement to the isotopic level.

Recent isotopic determinations of heavy neutron-capture elements
in metal-poor stars have yielded new insights on the roles of the
r- and s- processes nucleosynthesis in the early galaxy. The star
HD 175305 is a metal-poor star that has observed isotopic
fractions of the neutron-capture elements. The neutron-capture
elemental abundance pattern and isotopic fraction of this star can
be explained by a star formed in a molecular cloud that had been
polluted by both s- and r-process material. Overall the number of
stars with measured neutron-capture isotopic fractions is still
small. To underpin these studies, accurate abundance analysis for
similar metal-poor stars are required. Obviously, it is needed
that a more precise knowledge of the neutron-capture isotopic
composition of these stars. Further isotopic studies of these
stars will reveal the characteristics of the s- and r-processes at
low metallicity, such as stellar nucleosynthesis and the history
of enrichment of neutron-capture elements in the early galaxy.

\section*{Acknowledgments}
We thank the referee for an extensive and helpful review, containing
very relevant scientific advice that improved this paper greatly.
This work has been Supported by the National Natural Science
Foundation of China under Nos 10673002, 10973006, 10847119 and
10778616, the Natural Science Foundation of Hebei Provincial
Education Department under Grant No 2008127, the Science Foundation
of Hebei Normal University under Grant No L2007B07 and the Natural
Science Foundation of Hebei Proince under Grant no. A2009000251.

 \bsp

\label{lastpage}

\end{document}